\documentclass[12pt]{article}
\usepackage{sw20lart}

\input tcilatex
\begin{document}

\begin{center}
\textbf{PROOF\ OF\ SERBAN'S\ CONJECTURE}
\end{center}

\vskip 1.5cm\ 

\begin{center}
\textbf{M.C.Berg\`{e}re*}

\vskip 0.5cm

CEA-Saclay, Service de Physique Th\'{e}orique

F-91191 Gif sur Yvette Cedex, FRANCE
\end{center}

\vskip 1.5cm

\ 

\begin{center}
ABSTRACT
\end{center}

We prove Serban's conjecture which simplifies greatly the expression of the
advanced single-particle Green function in the Calogero-Sutherland model.
The importance of proving this conjecture is that it reorganizes the form
factor in terms of two dimensional Coulomb gaz correlators and confirms the
possible existence of a bosonization procedure for this system.

\vfill 

\noindent \ Submitted to Journal of Mathematical Physics.

\noindent *Membre du C.N.R.S.

\newpage\ 

\section{Introduction.}

\bigskip\ 

The Calogero-Sutherland model Ref.1, attracted much of attention during the
last few years, due to its connection to random matrix theory, fractional
statistics in one dimension, conformal field theory, etc...This model is
exactly solvable and its solution is known to a large extent.\ The
corresponding Hamiltonian 
\begin{equation}
H=-\frac 12\sum_{i=1}^N\frac{d^2}{dx_i^2}+\beta \left( \beta -1\right) \frac{%
\pi ^2}{L^2}\sum_{i<j}\frac 1{\sin ^2\left( \frac \pi L\left( x_i-x_j\right)
\right) }  \tag{1.1}
\end{equation}

\noindent describes the motion of N particles along a circle of length L.
The eigenstates of this Hamiltonian can be expressed Ref.2, in terms of the
so-called Jack polynomials in the variables $z_i=\exp (2i\pi x_i/L)$. For $%
\beta =0$ or 1, the particles behave like free spinless bosons or fermions,
and for rational $\beta $ the model describes particles with a fractional
statistic (excitations from the Fermi sea with kq quasi-particles and kp
quasi-holes if k is a positive integer and $\beta =p/q)$.

The two points correlation functions were calculated using two different
methods.\ The first one generalizes Dyson's work on the correlations of the
eigenvalues of random matrices Ref.3, which led to explicit expressions for
the static density-density correlations for $\beta $ =1/2,1,2. For these
values of $\beta ,$ Simons, Lee and Al'tshuler in Ref.3, using
supersymmetric techniques initiated by Efetov Ref.3, calculated the
dynamical density-density correlations; then Haldane and Zirnbauer obtained
in Ref.4, the retarded (for $\beta $=2) and the advanced (for $\beta $%
=1/2,1,2) single-particle Green functions.

The other method works for any rational $\beta $ and uses the properties of
Jack polynomials.\ The dynamical density-density correlations and the
retarded single-particle Green functions were obtained for integer $\beta $
by Lesage, Pasquier and Serban in Ref.5 and for rational $\beta $ by Ha in
Ref.6. The correlations functions can be expressed in terms of ''form
factors'' for some operator $A$: 
\begin{equation}
\left| F\right| ^2=\left| <v_1,..,v_p;w_1,..,w_q\mid A\mid 0>\right| ^2 
\tag{1.2}
\end{equation}

\noindent where the $v$'s and the $w$'s are the rapidities of the
quasi-holes and of the quasi-particles respectively. $F$ can be decomposed
as a product 
\[
F=F^{\left( 0\right) }\times F_\beta ^{\left( 1\right) } 
\]

\noindent The expression $F^{\left( 0\right) }$ is a purely statistical
contribution given by 
\begin{equation}
\left| F^{\left( 0\right) }\right| ^2=\frac{\prod_{i<j}\left| v_i-v_j\right|
^{2/\beta }\prod_{k<l}\left| w_k-w_l\right| ^{2\beta }}{\prod_i\left(
v_F^2-v_i^2\right) ^{1-1/\beta }\prod_k\left( w_k^2-v_F^2\right) ^{1-\beta
}\prod_{i,k}\left( v_i-w_k\right) ^2}  \tag{1.3}
\end{equation}

\noindent where $v_F$ is half the size (in rapidity) of the Fermi sea. In
the case of the advanced Green function ($A=\Psi ^{+})$ it was found in
Ref.4 that there are two contributions: one with one quasi-particle (zero
quasi-hole) and one with q+1 quasi-particles and p quasi-holes (if $\beta $%
=p/q).\ For this second contribution, $F_\beta ^{\left( 1\right) }$ was
found in Ref.4, for $\beta =2,1,1/2$, to be 
\begin{eqnarray}
F_2^{\left( 1\right) } &=&C_2\frac d{dz}\left( \frac{\left( z-v_1\right)
\left( z-v_2\right) }{\left( z-w_1\right) ^2}\right) _{z=w_0}  \nonumber \\
F_1^{\left( 1\right) } &=&0  \nonumber \\
F_{1/2}^{\left( 1\right) } &=&C_{1/2}\int_{w_0}^\infty \frac{dz}{\sqrt{z-w_0}%
}\left( \frac{z-v_1}{\sqrt{\left( z-w_1\right) \left( z-w_2\right) }}%
-1\right)  \tag{1.4}
\end{eqnarray}

Moreover, the authors proposed the following conjecture: for any positive
integer $\beta =p,$%
\begin{equation}
F_p^{\left( 1\right) }=C_p\left( \frac d{dz}\right) ^{p-1}\left\{ \frac{%
\prod_{i=1}^p\left( z-v_i\right) }{\left( z-w_1\right) ^p}\right\} _{z=w_0} 
\tag{1.5}
\end{equation}

\bigskip\ 

\bigskip\ 

Recently, the single particle advanced Green function have been calculated
for all rational $\beta $ by Serban, Lesage and Pasquier Ref.7. Their method
was to sum the corresponding combination of Jack polynomials over all Young
tableau with r ($w_0\;$in the thermodynamic limit) columns of length N-1, p
legs and q arms, and to take the thermodynamic limit, that is N going to
infinity. They obtained (1.3) for $F^{\left( 0\right) }$, and $F_\beta
^{\left( 1\right) }$ as 
\begin{eqnarray}
F_\beta ^{\left( 1\right) } &=&\frac{\prod_{i=0}^q\prod_{j=1}^p\left(
w_i-v_j\right) ^\beta }{\prod_{i<j}\left( v_i-v_j\right) \prod_{0\leq
i<j\leq q}\left( w_i-w_j\right) ^{2\beta -1}}.  \nonumber \\
&&.\int_{w_q}^{w_{q-1}}d\xi _{q-1}..\int_{w_2}^{w_1}d\xi _1\prod_{i<j}\left(
\xi _i-\xi _j\right) \prod_{i=1}^{q-1}\prod_{j=0}^q\left( \xi _i-w_j\right)
^{\beta -1}.  \nonumber \\
&&.\partial _{v_1}..\partial _{v_p}\left[ \prod_{i<j}\left( v_i-v_j\right)
\prod_{j=1}^p\left( \prod_{i=0}^q\left( w_i-v_j\right) ^{1-\beta
}\prod_{i=1}^{q-1}\left( \xi _i-v_j\right) ^{-1}\right) \right]  \tag{1.6}
\end{eqnarray}

\bigskip\ 

Then, Serban conjectured that this expression reduces to 
\begin{equation}
F_\beta ^{\left( 1\right) }=\left( \beta -1\right) \frac{\left[ \Gamma
\left( \beta \right) \right] ^q}{2i\pi }\oint_{C_w}dz\frac{%
\prod_{i=1}^p\left( v_i-z\right) }{\prod_{j=0}^q\left( w_j-z\right) ^\beta }
\tag{1.7}
\end{equation}

\noindent where the contour $C_w$ surrounds the points $w_1,..,w_q.$

The conjecture was proven for $\beta $ integer (that is Haldane and
Zirnbauer conjecture (1.5)), and the result (1.4) for $\beta =1/2$ \ has
been obtained from (1.7), as well as the numerical equality between (1.6)
and (1.7) for $\beta =$1/2 and 3/2.

\bigskip\ 

\bigskip\ 

In this publication, we prove Serban's conjecture, not only for $\beta =p/q,$
but for a set of different $\beta _i$ attached to each rapidity $w_i.\;$The
importance of proving this conjecture is due to the interpretation of the
result in terms of basic conformal operators. It was already pointed out by
Khveshchenko Ref.8, that the density-density correlation function and that
the one particle retarded Green function could be simply reexpressed in
terms of 2D Coulomb gaz correlators 
\begin{equation}
<V_{a_1}\left( z_1\right) ...V_{a_n}\left( z_n\right) >=\prod_{i<j}\left(
z_i-z_j\right) ^{a_ia_j}\;if\;\sum_{i=1}^na_i=0  \tag{1.8}
\end{equation}

\noindent and zero otherwise. Equation (1.7) proves, in the more complicate
situation of the advanced Green function, that the same property is true: 
\begin{eqnarray}
F &=&F^{\left( 0\right) }\times F_\beta ^{\left( 1\right) }=\frac{\left(
\beta -1\right) \left[ \Gamma \left( \beta \right) \right] ^q}{\prod_i\left(
v_F^2-v_i^2\right) ^{\frac{\beta -1}2}\prod_k\left( w_k^2-v_F^2\right) ^{%
\frac{1-\beta }2}}.  \nonumber \\
.\,\frac 1{2i\pi }\oint_{C_w}dz &<&V_{-\sqrt{\beta }}\left( z\right)
\;\prod_{i=1}^pV_{-1/\sqrt{\beta }}\left( v_i\right) \;\prod_{k=0}^qV_{\sqrt{%
\beta }}\left( w_k\right) \;>  \tag{1.9}
\end{eqnarray}
\noindent where a charge $\sqrt{\beta }\;$is given to the quasi-particule
operator, $-1/\sqrt{\beta }$ to the quasi-hole operator, and $-\sqrt{\beta }$
to the screening operator introduced in order to have conservation of the
charge. At the present time, we do not know what has to be done with the
terms coming from the Fermi sea ($v_F$ dependent) but this might be overcome
naturally if we later develop a bozonisation procedure of the
Calogero-Sutherland model. Several arguments plead in favour of such a
construction and specially in Ref.9, it is shown that the Jack polynomials
are the singular vectors of the Virasoro and of the $W_N$ algebra. It is
however surprising to see in (1.9) that the correlators $V_a$ are not taken
in complexified space-time points but in the rapidities. Let us finally
mention that the relation between (1.6) and (1.7) can certainly be
q-deformed if we work with the relativistic Ruijsenaars-Schneider model
Ref.10, where the Jack polynomials are q-deformed into the Macdonald
polynomials Ref.11.\ 

\bigskip\ 

\smallskip 

\section{Proof of the conjecture.\ }

Given a set of (q+1) variables $\beta _i$ (large enough to make all
integrals convergent), and two sets of variables, (q+1) variables $w_i$ and
p variables $v_i$, we define $C$ as

\begin{eqnarray}
C &=&\ \int_{w_q}^{w_{q-1}}d\xi _{q-1}...\int_{w_2}^{w_1}d\xi
_1\prod_{i<j}(\xi _i-\xi _j)\prod_{i=1}^{q-1}\prod_{j=0}^q\left| \xi
_i-w_j\right| ^{\beta _j-1}.  \nonumber \\
&&.\frac{\partial _{v_1}..\partial _{v_p}\left[
\prod_{j=0}^q\prod_{k=1}^p\left( w_j-v_k\right) ^{1-\beta
_j}\prod_{i=1}^{q-1}\prod_{k=1}^p\left( \xi _i-v_k\right)
^{-1}\prod_{k<l}\left( v_k-v_l\right) \right] }{\prod_{j=0}^q\prod_{k=1}^p%
\left( w_j-v_k\right) ^{1-\beta _j}\prod_{k<l}\left( v_k-v_l\right) } 
\nonumber \\
&&\,  \tag{2.1}
\end{eqnarray}

Our result is 
\begin{eqnarray}
C &=&\frac{\prod_{j=0}^q\Gamma \left( \beta _j\right) }{\Gamma \left(
\sum_{j=0}^q\beta _j-p-1\right) }\;\frac{\prod_{0\leq i<j\leq q}\left(
w_i-w_j\right) ^{\beta _i+\beta _j-1}}{\prod_{j=0}^q\prod_{k=1}^p\left(
w_j-v_k\right) }.  \nonumber \\
&&.\frac{(-1)}{2i\pi }\oint_{C_0}dz\frac{\prod_{k=1}^p\left( z-v_k\right) }{%
\left( w_0-z\right) ^{\beta _0}\prod_{j=1}^q\left( z-w_j\right) ^{\beta _j}}
\nonumber \\
\, &&  \tag{2.2}
\end{eqnarray}

\noindent where $w_0>w_1\;$and $C_0$ surrounds the cut $\left[ w_0,+\infty
\right] $ ; we suppose that $\sum_{j=0}^q\beta _j>p+1$ to ensure the
convergence of the integral in (2.2) (otherwise, subtractions should have to
be considered).

We note that if all the variables $\beta _j$ are equal to p/q, we have 
\begin{equation}
\frac{\prod_{j=0}^q\Gamma \left( \beta _j\right) }{\Gamma \left(
\sum_{j=0}^q\beta _j-p-1\right) }=\left( \beta -1\right) \left[ \Gamma
\left( \beta \right) \right] ^q  \tag{2.3}
\end{equation}

\bigskip\ \bigskip

\bigskip\ 

\ The proof of this result requires several steps of different nature; in
order to have an easier reading, the main part of the proof is written in
this section while four appendices are devoted to technical details.

The expression $C$ in (2.1) is made of two parts: on the first line, the
integrals d$\xi $, on the second line the derivatives $\partial _v$, the
factorization being prevented by the $\left( \xi -v\right) ^{-1}\;$terms. A
first appendix (A) generalizes Cauchy's identity in order to eliminate the $%
\left( \xi -v\right) ^{-1}$terms. $C$ becomes a sum of terms, each of them
being factorized into d$\xi $ integrals and $\partial _v$ derivatives
without ($\xi -v$)$^{-1}$ terms. The second appendix (B) generalizes a
result by Dixon (1905) Ref.12, who performs a change of variables which
makes obvious the calculation of the integrals $d\xi $; the generalization
consists in adding the variable $w_0$ to Dixon's result which generates a
contour integral $\oint dz$. These two appendices are sufficient to prove
the conjecture in the case (q-1)$\geq $p. However, the case (q-1)$\leq p$ is
more difficult because we have extra $\partial _v$ derivatives to perform
over terms which do not contain $\left( \xi -v\right) ^{-1}$ anymore.
Appendix C is devoted to the calculation and to the properties of these $%
\partial _v$ derivatives. Finally, appendix D, which also, is only needed in
the case q-1$\leq $p, develops the properties of the integrals 
\begin{equation}
I_P=\frac 1{2i\pi }\oint_{C_0}dz\frac{P(z)}{\left( w_0-z\right) ^{\beta
_0}\prod_{j=1}^q\left( z-w_j\right) ^{\beta _j}}  \tag{2.4}
\end{equation}

\noindent where P(z) is a polynomial in z and where $C_0$ surrounds the cut $%
\left[ w_0,+\infty \right] $.

Let us mention that the proof is written for a set of generic values of the
variables $v_i$ without care to the case where several $v$'s are equal;
however, our result remain valid in that case since (2.1) has no pole
singularity at $v_i=v_j$ (AppendixC (C2)).

\bigskip\ \bigskip\ \bigskip\ 

\bigskip\ \bigskip

\bigskip\ \bigskip\ \bigskip\ \underline{First case:q-1$\geq $p.}

In expression (2.1), we use the generalized Cauchy's identity (A3) to write 
\begin{equation}
\frac{\prod_{i<j}\left( \xi _i-\xi _j\right) \prod_{k<l}\left(
v_k-v_l\right) }{\prod_{i=1}^{q-1}\prod_{j=1}^p\left( \xi _i-v_j\right) }%
=\sum_{I_p}\left( -\right) ^{\delta _{I_p}}\;\det \left| \frac 1{\xi
_{j_k}-v_j}\right| \;\prod\Sb i<j  \\ i,j\notin I_p  \endSb \left( \xi
_i-\xi _j\right)  \tag{2.5}
\end{equation}

\noindent where $I_p=(j_1<j_2<..<j_p)\subset (1,..,q-1)$ and $\delta
_{I_p}=(q-1)p-\sum_{i=1}^pj_i.$

\noindent Because of the absence of terms of the type $\left( v_i-v_j\right)
\;$on the right hand side of (2.5), and because each term of the determinant
is a product of $\left( \xi -v_j\right) ^{-1}$ for different j's, the action
of the derivatives $\partial _{v_j}$ in (2.1) factorizes. We get a product
of the type 
\begin{eqnarray}
&&\prod_{j=1}^p\frac{\partial _{v_j}\left[ \prod_{i=0}^q\left(
w_i-v_j\right) ^{1-\beta _i}\left( \xi _{a_j}-v_j\right) ^{-1}\right] }{%
\prod_{i=0}^q\left( w_i-v_j\right) ^{1-\beta _i}}  \nonumber \\
&=&\prod_{j=1}^p\left\{ \left[ \sum_{i=0}^q\frac{\beta _i-1}{w_i-v_j}-\frac
\partial {\partial \xi _{a_j}}\right] \left( \xi _{a_j}-v_j\right)
^{-1}\right\}  \tag{2.6}
\end{eqnarray}
\noindent where the $a_j$'s are a permutation of the $j_k$'s in $I_p$.The
right hand side of (2.6) can now be taken inside the integrals $d\xi $ to be
integrated by parts (we assume that the corresponding Re$\beta _i$'s 
\TEXTsymbol{>}1 to avoid the contribution of the end-points); the expression
(2.6) becomes 
\begin{equation}
\prod_{j=1}^p\left[ \sum_{i=0}^q\frac{\beta _i-1}{\left( w_i-v_j\right)
\left( \xi _{a_j}-w_i\right) }\right]  \tag{2.7}
\end{equation}

\noindent For any product of terms in the determinant (2.5), we have
obtained a product of terms (2.7); consequently, the determinant in (2.5)
can be replaced by the determinant 
\begin{equation}
\det \left| \sum_{i=0}^q\frac{\beta _i-1}{\left( \xi _{j_k}-w_i\right)
\left( w_i-v_j\right) }\right| =\sum_{J_p}\prod_{i\in J_p}\left( \beta
_i-1\right) \;\det \left| \frac 1{\xi _{j_k}-w_{l_j}}\right| \;\det \left|
\frac 1{w_{l_k}-v_j}\right|  \tag{2.8}
\end{equation}

\noindent where $J_p=(l_1<l_2<..<l_p)\subset (0,..,q).$

\bigskip\ 

The situation now, is that the determinant $\left| \left( \xi
_{j_k}-v_i\right) ^{-1}\right| \;$in (2.5) is replaced by the determinant$%
\;\left| \left( \xi _{j_k}-w_{l_j}\right) ^{-1}\right| \;$with the $l_j$'s
in $J_p$. We may now use the generalized Cauchy's identity (A3) backwards to
sum over the different $I_p$'s and we obtain, inside the sum over $J_p$, 
\begin{equation}
\frac{\prod_{i<j}\left( \xi _i-\xi _j\right) \prod_{j<k}\left(
w_{l_j}-w_{l_k}\right) }{\prod_{i=1}^{q-1}\prod_{k=1}^p\left( \xi
_i-w_{l_k}\right) }=\left( -\right) ^{\delta _{J_p}-\eta }\;\frac{%
\prod_{i<j}\left( \xi _i-\xi _j\right) \prod_{j<k}\left(
w_{l_j}-w_{l_k}\right) }{\prod_{i=1}^q\prod_{k=1}^p\left| \xi
_i-w_{l_k}\right| }  \tag{2.9}
\end{equation}

\noindent where the $l_j$'s belong to $J_p$, and $\delta
_{J_p}=pq-\sum_{k=1}^pl_k$ (here, $\eta =$ 1 if $l_1=0$ and 0 otherwise).

\bigskip\ 

\bigskip\ 

We proved, up to now, that 
\begin{eqnarray}
C &=&\frac 1{\prod_{k<l}\left( v_k-v_l\right) }\sum_{J_p}\left( -\right)
^{\delta _{J_p}}\prod_{i\in J_p}\left( \beta _i-1\right) \prod_{j<k}\left(
w_{l_j}-w_{l_k}\right) \det \left| \frac 1{w_{l_k}-v_j}\right| .  \nonumber
\\
&&.\int_{w_q}^{w_{q-1}}d\xi _{q-1}...\int_{w_2}^{w_1}d\xi _1\prod_{i<j}(\xi
_i-\xi _j)\;\prod_{i=1}^{q-1}\prod_{j=0}^q\left| \xi _i-w_j\right| ^{\beta
_j^{\prime }-1}  \tag{2.10}
\end{eqnarray}

\noindent where $\beta _j^{\prime }=\beta _j$ if $j\notin J_p$ and $\beta
_j^{\prime }=\beta _j-1\;$if $j\in J_p.$

\bigskip\ 

\bigskip\ 

We are now in situation to use Dixon's result with an extra variable $w_0$
(B4,B10) and to write the second line of (2.10) as 
\begin{eqnarray}
&&\frac{\prod_{j=0}^q\Gamma \left( \beta _j^{\prime }\right) }{\Gamma \left(
\sum_{j=0}^q\beta _j-p-1\right) }\prod_{0\leq i<j\leq q}\left(
w_i-w_j\right) ^{\beta _i^{\prime }+\beta _j^{\prime }-1}.  \nonumber \\
&&.\frac{(-1)}{2i\pi }\oint_{C_0}dz\frac 1{\left( w_0-z\right) ^{\beta
_0^{\prime }}\prod_{j=1}^q\left( z-w_j\right) ^{\beta _j^{\prime }}} 
\tag{2.11}
\end{eqnarray}

\noindent The expression $C$ becomes 
\begin{eqnarray}
C &=&\frac{\prod_{j=0}^q\Gamma \left( \beta _j\right) }{\Gamma \left(
\sum_{j=0}^q\beta _j-p-1\right) }\frac{\prod_{0\leq i<j\leq q}\left(
w_i-w_j\right) ^{\beta _i+\beta _j-2}}{\prod_{k<l}\left( v_k-v_l\right) }. 
\nonumber \\
&&.\frac{(-1)}{2i\pi }\oint_{C_0}dz\frac 1{\left( w_0-z\right) ^{\beta
_0}\prod_{j=1}^q\left( z-w_j\right) ^{\beta _j}}.  \nonumber \\
&&.\sum_{J_p}\left( -\right) ^{\delta _{J_p}}\prod_{k=1}^p\left(
z-w_{l_k}\right) \det \left| \frac 1{w_{l_k}-v_j}\right| \prod\Sb i<j  \\ %
i,j\notin J_p  \endSb \left( w_i-w_j\right)  \tag{2.12}
\end{eqnarray}

$\;\;\;\;$\noindent Again, as a consequence of the generalized Cauchy's
identity, we may now sum over all $J_p$'s; from (A8) with q+1\TEXTsymbol{>}%
p; the third line of (2.12) is nothing but 
\begin{equation}
\prod_{k=1}^p\left( z-v_k\right) \frac{\prod_{0\leq i<j\leq q}\left(
w_i-w_j\right) \prod_{k<l}\left( v_k-v_l\right) }{\prod_{i=0}^q\prod_{j=1}^p%
\left( w_i-v_j\right) }  \tag{2.13}
\end{equation}

\noindent The fact that the indices of the $w$'s run from 0 to q (instead of
1 to q+1) makes in (A8), a shift by p of $\delta _{J_p}$; this explains why
we do not have a $\left( -\right) ^p$ in (2.13).

This result ends the proof for the case (q-1)$\geq $p.

\bigskip\ 

\bigskip\ 

\bigskip\ 

\ \ \underline{Second case: q-1$\leq p.$}

\bigskip\ Again, we use the generalized Cauchy's identity in order to
calculate the left hand side of (2.5), but here we have more variables $v_j$
than variables $\xi _i$; using the relation (A2), we write 
\begin{equation}
\frac{\prod_{i<j}\left( \xi _i-\xi _j\right) \prod_{k<l}\left(
v_k-v_l\right) }{\prod_{i=1}^{q-1}\prod_{j=1}^p\left( \xi _i-v_j\right) }%
=\sum_{I_{q-1}}\left( -\right) ^{\delta _{I_{q-1}}}\det \left| \frac 1{\xi
_i-v_{j_k}}\right| \prod\Sb k<l  \\ k,l\notin I_{q-1}  \endSb \left(
v_k-v_l\right)  \tag{2.14}
\end{equation}

\noindent where $I_{q-1}=(j_1<j_2<...<j_{q-1})\subset \left( 1,...,p\right)
\;$and $\delta _{I_{q-1}}=\sum_{k=1}^{q-1}j_k-q+1.$ Consequently, the
derivatives $\partial _{v_j}$ split into two parts depending whether j is in 
$I_{q-1}$ or not. The second line of (2.1) becomes 
\begin{eqnarray}
\;\;\; &&\frac 1{\prod_{k<l}\;(v_k-v_l)}\sum_{I_{q-1}}\left( -\right)
^{\delta _{I_{q-1}}}.  \nonumber \\
&&.\frac{\prod_{j\notin I_{q-1}}(\partial _{v_j})\left[
\prod_{i=0}^q\prod_{j\notin I_{q-1}}\left( w_i-v_j\right) ^{1-\beta
_i}\prod_{\Sb k<l  \\ k,l\notin I_{q-1}  \endSb }\left( v_k-v_l\right)
\right] }{\prod_{i=0}^q\prod_{j\notin I_{q-1}}\left( w_i-v_j\right)
^{1-\beta _i}}.  \nonumber \\
&&.\frac{\prod_{j\in I_{q-1}}\left( \partial _{v_j}\right) \left[
\prod_{i=0}^q\prod_{j\in I_{q-1}}\left( w_i-v_j\right) ^{1-\beta _i}\det
\left| \frac 1{\xi _i-v_{j_k}}\right| \right] }{\prod_{i=0}^q\prod_{j\in
I_{q-1}}\left( w_i-v_j\right) ^{1-\beta _i}}  \tag{2.15}
\end{eqnarray}

In the third line of (2.15), the derivatives $\partial _v$ factorize so that
this term can be calculated in exactly the same way as for the case (q-1)$%
\geq p.$ The determinant $\left( \xi _i-v_{j_k}\right) ^{-1}$ can be
replaced in the same manner by 
\begin{equation}
\det \left| \sum_{r=0}^q\frac{\beta _r-1}{\left( \xi _i-w_r\right) \left(
w_r-v_{j_k}\right) }\right| =\sum_{J_{q-1}}\;\prod_{r\in J_{q-1}}\left(
\beta _r-1\right) \;\det \left| \frac 1{\xi _i-w_r}\right| \det \left| \frac
1{w_r-v_{j_k}}\right|  \tag{2.16}
\end{equation}

\noindent where $J_{q-1}=(l_1<l_2<...<l_{q-1})\subset (0,...,q).$ We now use
Cauchy's identity (A1), to write 
\begin{equation}
\det \left| \frac 1{\xi _i-w_r}\right| =\left( -\right) ^{\frac{q\left(
q-1\right) }2+\delta _{J_{q-1}}-\eta }\;\frac{\prod_{i<j}(\xi _i-\xi
_j)\prod_{j<k}\left( w_{l_j}-w_{l_k}\right) }{\prod_{i=1}^{q-1}%
\prod_{j=1}^{q-1}\left| \xi _i-w_{l_j}\right| }  \tag{2.17}
\end{equation}

\noindent where $\delta _{J_{q-1}}=\sum_{j=1}^{q-1}l_j-q+1$ and $\eta =1$ if 
$l_1=0$ (and 0 otherwise).

\bigskip\ 

Again, after inserting (2.17) into the integrals d$\xi $, we may integrate
using Dixon's change of variables (B4, B10) to get 
\begin{eqnarray}
&&\left( -\right) ^{\frac{q\left( q-1\right) }2}.\frac{\prod_{j=0}^q\Gamma
\left( \beta _j\right) }{\Gamma \left( \sum_{j=0}^q\beta _j-q\right) }%
\prod_{0\leq i<j\leq q}\left( w_i-w_j\right) ^{\beta _i+\beta _j-2}. 
\nonumber \\
&&.\frac{(-1)}{2i\pi }\oint_{C_0}dz\frac 1{\left( w_0-z\right) ^{\beta
_0}\prod_{j=1}^q\left( z-w_j\right) ^{\beta _j}}.  \nonumber \\
&&.\sum_{J_{q-1}}\left( -\right) ^{\delta _{J_{q-1}}}\prod_{j=1}^{q-1}\left(
z-w_{l_j}\right) \det \left| \frac 1{w_{l_j}-v_{j_k}}\right| \prod\Sb i<j 
\\ i,j\notin J_{q-1}  \endSb \left( w_i-w_j\right)  \tag{2.18}
\end{eqnarray}

\noindent Using a consequence of the generalized Cauchy's identity for a
given set of q-1 variables $v_{j_k}$'s in $I_{q-1}$, the third line of
(2.18) can be summed over $J_{q-1}$; here again, because the $w$'s are
labelled from 0 to q+1, $\delta _{Jq-1}$ has to be shifted by q-1 in order
to apply (A8) with q+1\TEXTsymbol{>}q-1. We get for the third line of (2.15) 
\begin{eqnarray}
&&\left( -\right) ^{\frac{\left( q-1\right) \left( q-2\right) }2}\frac{%
\prod_{j=0}^q\Gamma \left( \beta _j\right) }{\Gamma \left( \sum_{j=0}^q\beta
_j-q\right) }\frac{\prod_{0\leq i<j\leq q}\left( w_i-w_j\right) ^{\beta
_i+\beta _j-1}\prod\Sb k<l  \\ k,l\in I_{q-1}  \endSb \left( v_k-v_l\right) 
}{\prod_{i=0}^q\prod_{j\in I_{q-1}}\left( w_i-v_j\right) }.  \nonumber \\
&&.\frac{(-1)}{2i\pi }\oint_{C_0}dz\frac{\prod_{j\in I_{q-1}}\left(
z-v_j\right) }{\left( w_0-z\right) ^{\beta _0}\prod_{j=1}^q\left(
z-w_j\right) ^{\beta _j}}  \tag{2.19}
\end{eqnarray}

\bigskip\ 

\bigskip\ 

At this point, we proved that, in the case (q-1)$\leq $p, the expression $C$
may be written as 
\begin{eqnarray}
C &=&\left( -\right) ^{\frac{\left( q-1\right) \left( q-2\right) }2}\frac{%
\prod_{j=0}^q\Gamma \left( \beta _j\right) }{\Gamma \left( \sum_{j=0}^q\beta
_j-q\right) }\frac{\prod_{0\leq i<j\leq q}\left( w_i-w_j\right) ^{\beta
_i+\beta _j-1}}{\prod_{i=0}^q\prod_{j=1}^p\left( w_i-v_j\right)
\prod_{k<l}\left( v_k-v_l\right) }.  \nonumber \\
. &&\sum_{I_{q-1}}\left( -\right) ^{\delta _{I_{q-1}}}\frac{\prod_{j\notin
_{I_{q-1}}}\left( \partial _{v_j}\right) \left[ \prod_{i=0}^q\prod_{j\notin
I_{q-1}}\left( w_i-v_j\right) ^{1-\beta _i}\prod\Sb k<l  \\ k,l\notin
I_{q-1}  \endSb \left( v_k-v_l\right) \right] }{\prod_{i=0}^q\prod_{j\notin
I_{q-1}}\left( w_i-v_j\right) ^{-\beta _i}}.  \nonumber \\
. &&\prod\Sb k<l  \\ k,l\in I_{q-1}  \endSb \left( v_k-v_l\right) \frac{(-1)%
}{2i\pi }\oint_{C_0}dz\frac{\prod_{j\in I_{q-1}}\left( z-v_j\right) }{\left(
w_0-z\right) ^{\beta _0}\prod_{j=1}^q\left( z-w_j\right) ^{\beta _j}} 
\tag{2.20}
\end{eqnarray}

\bigskip\ 

The purpose of appendices C and D is to calculate the sum over $I_{q-1}$ in
(2.20). In (D23) we show that the second and third lines of (2.20) are equal
to 
\begin{eqnarray}
&&\left( -\right) ^{\frac{\left( q-1\right) \left( q-2\right) }2}\frac{%
\Gamma \left( \sum_{j=0}^q\beta _j-q\right) }{\Gamma \left(
\sum_{j=0}^q\beta _j-p-1\right) }\prod_{k<l}\left( v_k-v_l\right) . 
\nonumber \\
&&.\frac{\left( -1\right) }{2i\pi }\oint_{C_0}dz\frac{\prod_{k=1}^p\left(
z-v_k\right) }{\left( w_0-z\right) ^{\beta _0}\prod_{j=1}^q\left(
z-w_j\right) ^{\beta _j}}  \tag{2.21}
\end{eqnarray}

\noindent and this ends the proof of the conjecture in the case (q-1)$\leq $%
p.

\bigskip\ 

\bigskip\ 

\bigskip\ 

\bigskip

\begin{center}
\textbf{ACKNOWLEDGMENTS}

\ 
\end{center}

I thank D.Serban for her constant encouragement, for many discussions about
the work in Ref.7 and for her careful reading of the manuscript.

\bigskip\ 

\bigskip\ 

\bigskip

\newpage\ 

\section{Appendix A}

\bigskip\ 

\bigskip\ 

Given two sets of variables ($x_1,...,x_n)$ and $\left( y_1,...,y_m\right) $,%
$\;$if n=m, Cauchy's identity is 
\begin{equation}
\frac{\prod_{i<j}\left( x_i-x_j\right) \prod_{k<l}\left( y_k-y_l\right) }{%
\prod_{i=1}^n\prod_{k=1}^n\left( x_i-y_k\right) }=\left( -\right) ^{\frac{%
n\left( n-1\right) }2}\det \left| \frac 1{x_i-y_k}\right|  \tag{A1}
\end{equation}

\noindent (apart from the sign, this result is relatively evident from the
properties of the determinant, the homogeneity and the pole structure). Now,
if n$\leq $m, we may generalize this relation by systematically organizing
the residues of the poles; we get: 
\begin{equation}
\frac{\prod_{i<j}\left( x_i-x_j\right) \prod_{k<l}\left( y_k-y_l\right) }{%
\prod_{i=1}^n\prod_{k=1}^m\left( x_i-y_k\right) }=\sum_{I_n}\left( -\right)
^{\delta _{I_n}}\det \left| \frac 1{x_i-y_{j_k}}\right| \prod\Sb k<l  \\ %
k,l\notin I_n  \endSb \left( y_k-y_l\right)  \tag{A2}
\end{equation}

\noindent where $I_n=\left( j_1<j_2<...<j_n\right) \subset \left(
1,...,m\right) $ and $\delta _{I_n}=\sum_{i=1}^nj_i-n.$ Of course, if n$\geq
m,$ the situation is symmetric and we have 
\begin{equation}
\frac{\prod_{i<j}\left( x_i-x_j\right) \prod_{k<l}\left( y_k-y_l\right) }{%
\prod_{i=1}^n\prod_{k=1}^m\left( x_i-y_k\right) }=\sum_{I_m}\left( -\right)
^{\delta _{I_m}}\det \left| \frac 1{x_{l_i}-y_k}\right| \prod\Sb i<j  \\ %
i,j\notin I_m  \endSb \left( x_i-x_j\right)  \tag{A3}
\end{equation}

\noindent where $I_m=\left( l_1<l_2<...<l_m\right) \subset \left(
1,...,n\right) $ and $\delta _{I_m}=mn-\sum_{k=1}^ml_k.$ We call (A2-3)
generalized Cauchy's identities. In (A2-3), we adopt the convention that $%
\prod\Sb i<j  \\ i,j\notin I_{.}  \endSb (z_i-z_j)=1$ if there is 0 or 1
variable $z$ $\notin I_{.}$.

Let us note the following property: for n\TEXTsymbol{<}m, if we let a given
variable $x_p\rightarrow \infty $, the left-hand side of (A2) behaves as $%
x_p^{-\left( m-n+1\right) }\leq x_p^{-2}$ while the right-hand side behaves
as $x_p^{-1}.$ Consequently, the coefficient of $x_p^{-1}$in the right-side
of (A2) is zero. This coefficient can be calculated: we define the matrix $%
\Delta _{I_n,p}$ such that 
\begin{eqnarray}
\left( \Delta _{I_n,p}\right) _{ik} &=&\frac 1{x_i-y_{j_k}}\;for\;i\neq p 
\tag{A4} \\
&=&1\;for\;i=p  \nonumber
\end{eqnarray}

\noindent then, 
\begin{equation}
\sum_{I_n}\left( -\right) ^{\delta _{I_n}}\det \left( \Delta _{I_n,p}\right)
\prod\Sb k<l  \\ k,l\notin I_n  \endSb \left( y_k-y_l\right) =0  \tag{A5}
\end{equation}

\bigskip\ 

We are now using (A5) to calculate 
\begin{equation}
\Phi \left( z\right) =\sum_{I_n}\left( -\right) ^{\delta
_{I_n}}\prod_{k=1}^n\left( z-y_{j_k}\right) \det \left| \frac
1{x_i-y_{j_k}}\right| \prod\Sb k<l  \\ k,l\notin I_n  \endSb \left(
y_k-y_l\right)  \tag{A6}
\end{equation}

\noindent The determinant in (A6) is a sum of terms of the type $\left(
x_1-y_{\alpha _1}\right) ^{-1}\left( x_2-y_{\alpha _2}\right) ^{-1}$

$..\left( x_n-y_{\alpha _n}\right) ^{-1}$ where $\left( \alpha _1,..,\alpha
_n\right) $ is a permutation of $\left( j_1,..,j_n\right) .$ For such a
term, we write 
\[
\prod_{k=1}^n\left( z-y_{j_k}\right) =\prod_{k=1}^n\left( z-y_{\alpha
_k}\right) =\sum_J\prod_{k\notin J}\left( z-x_k\right) \prod_{j\in J}\left(
x_j-y_{\alpha _j}\right) 
\]

\noindent where $J$ is a set of $\left| J\right| $ indices $\subset \left(
1,..,n\right) .$ Consequently, when summing over all terms of the
determinant, we get 
\[
\prod_{k=1}^n\left( z-y_{j_k}\right) \det \left| \frac 1{x_i-y_{j_k}}\right|
=\sum_J\prod_{k\notin J}\left( z-x_k\right) \det \left| \left( \frac
1{x_i-y_{j_k}}\right) _J\right| 
\]

\noindent where the matrices $\left( \left( x_i-y_{j_k}\right) ^{-1}\right)
_J$ have $\left| J\right| $ lines of 1 corresponding to the indices i $\in
J. $ Their determinants are trivially zero for $\left| J\right| \geq 2.$
Now, for $\left| J\right| =1,$ the sum over $I_n$ in (A6) gives zero because
of (A5); the only non vanishing contribution comes from $\left| J\right| =0.$
We just proved that 
\begin{equation}
\Phi \left( z\right) =\prod_{k=1}^n\left( z-x_k\right) \;\frac{%
\prod_{i<j}\left( x_i-x_j\right) \prod_{k<l}\left( y_k-y_l\right) }{%
\prod_{i=1}^n\prod_{j=1}^m\left( x_i-y_j\right) }  \tag{A7}
\end{equation}

\bigskip\ 

By symmetry, for n\TEXTsymbol{>}m, we have 
\begin{eqnarray}
&&\sum_{I_m}\left( -\right) ^{\delta _{I_m}}\prod_{i=1}^m\left(
z-x_{l_i}\right) \det \left| \frac 1{x_{l_i}-y_k}\right| \prod\Sb i<j  \\ %
i,j\notin I_m  \endSb \left( x_i-x_j\right)  \nonumber \\
&=&\prod_{i=1}^m\left( z-y_i\right) \;\frac{\prod_{i<j}\left( x_i-x_j\right)
\prod_{k<l}\left( y_k-y_l\right) }{\prod_{i=1}^n\prod_{j=1}^m\left(
x_i-y_j\right) }  \tag{A8}
\end{eqnarray}

\noindent where, this time, we used determinants of matrices with columns of
(-1)'s. Let us mention that if the indices of the $x$'s run from 0 to n-1
instead of 1 to n (as it is the case in eq.(2.12,2.18)), then, $\delta
_{I_m} $ has to be replaced in (A8) by $\delta _{I_m}+m.$

\bigskip\ 

\bigskip\ 

\section{Appendix B}

\bigskip\ 

In this appendix, we calculate the integrals $d\xi $ which appear in the
first line of (2.1). We first calculate the integrals without the variable $%
w_0;$ the end of the appendix is devoted to the introduction of $w_0$
generating a contour integral in the complex plane. Let us define 
\begin{equation}
J_{q-1}=\int_{w_q}^{w_{q-1}}d\xi _{q-1}...\int_{w_2}^{w_1}d\xi
_1\prod_{i<j}\left( \xi _i-\xi _j\right)
\prod_{i=1}^{q-1}\prod_{j=1}^q\left| \xi _i-w_j\right| ^{\beta _j-1}. 
\tag{B1}
\end{equation}

\bigskip\ 

Clearly, 
\[
J_1=\frac{\Gamma \left( \beta _1\right) \Gamma \left( \beta _2\right) }{%
\Gamma \left( \beta _1+\beta _2\right) }\;\left( w_1-w_2\right) ^{\beta
_1+\beta _2-1} 
\]

\noindent A naive calculation of $J_2$ gives three products of
hypergeometric functions, namely 
\begin{eqnarray}
J_2 &=&\prod_{i<j}\left( w_i-w_j\right) ^{\beta _i+\beta _j-1}\;\frac{\Gamma
\left( \beta _1\right) \left( \Gamma \left( \beta _2\right) \right) ^2\Gamma
\left( \beta _3\right) }{\Gamma \left( \beta _1+\beta _2\right) \Gamma
\left( \beta _2+\beta _3\right) }.  \nonumber \\
&&.\left[ 
\begin{array}{c}
\begin{array}{c}
\;\;F\left( \beta _3,-\beta _1;\beta _2+\beta _3;\alpha \right)
\,\,\,\;F\left( 1-\beta _3,\beta _1;\beta _1+\beta _2;1-\alpha \right)
\end{array}
\\ 
+F\left( \beta _3,1-\beta _1;\beta _2+\beta _3;\alpha \right) \;F\left(
-\beta _3,\beta _1;\beta _1+\beta _2;1-\alpha \right) \\ 
\;-F\left( \beta _3,1-\beta _1;\beta _2+\beta _3;\alpha \right) \;F\left(
1-\beta _3,\beta _1;\beta _1+\beta _2;1-\alpha \right)
\end{array}
\right]  \tag{B2}
\end{eqnarray}

\noindent where $\alpha =\left( w_2-w_3\right) /\left( w_1-w_3\right) $. The
remarkable fact is that the square bracket $\left[ ..\right] $ in (B2) is $%
\alpha $ independant and equal to 
\[
\frac{\Gamma \left( \beta _1+\beta _2\right) \;\Gamma \left( \beta _2+\beta
_3\right) }{\Gamma \left( \beta _1+\beta _2+\beta _3\right) \;\Gamma \left(
\beta _2\right) } 
\]

\noindent This result is due to Elliot Ref.13, who generalizes Legendre's
relation from the theory of elliptic integrals. A generalization of Elliot's
result has been given by Dixon Ref.12, who proved that 
\begin{equation}
J_{q-1}=\frac{\prod_{j=1}^q\Gamma \left( \beta _j\right) }{\Gamma \left(
\sum_{j=1}^q\beta _j\right) }\;\prod_{i<j}\left( w_i-w_j\right) ^{\beta
_i+\beta _j-1}  \tag{B3}
\end{equation}

\noindent by performing a clever change of variables: given the function 
\[
f\left( \theta \right) =\prod_{j=1}^q\left( \theta -w_j\right) 
\]

\noindent we define new variables 
\[
x_j=\left( -\right) ^{q-1}\frac{\prod_{i=1}^{q-1}\left( \xi _j-w_i\right) }{%
f^{\prime }\left( w_j\right) }\;for\;j=1,...,q 
\]

\noindent It is easy to verify that all variables $x_j\;$are $\geq 0$;
moreover, we have 
\[
\sum_{j=1}^qx_j=1 
\]

\noindent so that $x_q$ can be considered as dependant of the other $%
x^{\prime }s.$ The jacobian of the transformation is 
\[
\frac{d\left( \xi _1,..,\xi _{q-1}\right) }{d\left( x_1,...,x_{q-1}\right) }=%
\frac{\prod_{i<j}\left( w_i-w_j\right) }{\prod_{i<j}\left( \xi _i-\xi
_j\right) } 
\]

\noindent Consequently, 
\[
J_{q-1}=\prod_{i<j}\left( w_i-w_j\right) ^{\beta _i+\beta
_j-1}\int_0^1..\int_0^1\prod_{j=1}^q\left[ dx_j\;x_j^{\beta _j-1}\right]
\;\delta \left( \sum_{j=1}^qx_j-1\right) 
\]

\noindent which proves (B3).

\bigskip\ 

\bigskip\ 

Next, we wish to calculate (or to transform into a single contour integral
in the complex plane) the integrals 
\begin{equation}
K_{q-1}=\int_{w_q}^{w_{q-1}}d\xi _{q-1}...\int_{w_2}^{w_1}d\xi
_1\prod_{i<j}\left( \xi _i-\xi _j\right)
\prod_{i=1}^{q-1}\prod_{j=0}^q\left| \xi _i-w_j\right| ^{\beta _j-1} 
\tag{B4}
\end{equation}

In order to transform $K_{q-1}$, we write 
\begin{equation}
\prod_{i=1}^{q-1}\left| \xi _i-w_0\right|
=\sum_{j=1}^q\;A_j\;\prod_{i=1}^{q-1}\left| \xi _i-w_j\right|  \tag{B5}
\end{equation}

\noindent with 
\begin{equation}
A_j=\frac{\prod_{k\neq j}\left( w_0-w_k\right) }{\prod_{k\neq j}\left|
w_j-w_k\right| }  \tag{B6}
\end{equation}

\bigskip\ 

Let us first consider the case where $\beta _0=n$ integer.\ Then, 
\begin{equation}
\prod_{i=1}^{q-1}\left| \xi _i-w_0\right| ^{n-1}=\left( n-1\right) !\sum\Sb %
\left\{ p_j\right\}  \\ \sum_{j=1}^qp_j=n-1  \endSb \prod_{j=1}^q(\frac{%
A_j^{p_j}}{p_j!}\;\prod_{i=1}^{q-1}\left| \xi _i-w_j\right| ^{p_j})  \tag{B7}
\end{equation}

\noindent so that we can use (B1, B3) to integrate the $d\xi \;$integrals
with $\beta _j$ replaced by $\beta _j+p_j.$ We get 
\begin{eqnarray}
K_{q-1} &=&\frac{\left( n-1\right) !}{\Gamma \left( \sum_{j=1}^q\beta
_j+n-1\right) }\prod_{1\leq i<j\leq q}\left( w_i-w_j\right) ^{\beta _i+\beta
_j-1}.  \nonumber \\
&&.\sum\Sb \left\{ p_j\right\}  \\ \sum_{j=1}^qp_j=n-1  \endSb %
\prod_{j=1}^q\left\{ \frac{\Gamma \left( \beta _j+p_j\right) }{p_j!}\left(
w_0-w_j\right) ^{n-1-p_j}\right\}  \tag{B8}
\end{eqnarray}

On the other hand, the contour integral around$\;z=w_0$ 
\[
\frac 1{2i\pi }\oint_{C_0}dz\;\frac 1{\left( z-w_0\right)
^n\prod_{j=1}^q\left( z-w_j\right) ^{\beta _j}} 
\]

\noindent is a derivative 
\[
\frac 1{\left( n-1\right) !}\frac{\partial ^{n-1}}{\partial z^{n-1}}\left\{
\frac 1{\prod_{j=1}^q\left( z-w_j\right) ^{\beta _j}}\right\} _{z=w_0} 
\]

\noindent equal to 
\begin{eqnarray}
&&\frac{\left( -\right) ^{n-1}}{\prod_{j=1}^q\left[ \Gamma \left( \beta
_j\right) \left( w_0-w_j\right) ^{\beta _j+n-1}\right] }.  \nonumber \\
&&.\sum\Sb \left\{ p_j\right\}  \\ \sum_{j=1}^qp_j=n-1  \endSb %
\prod_{j=1}^q\left\{ \frac{\Gamma \left( \beta _j+p_j\right) }{p_j!}\left(
w_0-w_j\right) ^{n-1-p_j}\right\}  \tag{B9}
\end{eqnarray}

\noindent If we compare (B8) and (B9), we get for $\beta _0=n$ integer 
\begin{eqnarray}
K_{q-1} &=&\frac{\prod_{j=0}^q\Gamma \left( \beta _j\right) }{\Gamma \left(
\sum_{j=0}^q\beta _j-1\right) }\prod_{0\leq i<j\leq q}\left( w_i-w_j\right)
^{\beta _i+\beta _j-1}.  \nonumber \\
&&.\frac{(-1)}{2i\pi }\oint_{C_0}dz\frac 1{\left( w_0-z\right) ^{\beta
_0}\prod_{j=1}^q\left( z-w_j\right) ^{\beta _j}}  \tag{B10}
\end{eqnarray}

\bigskip\ 

\bigskip\ 

We now prove that (B10) is also true for any $\beta $'s such that the
integral $dz$ converges around a contour $C_0$ which surrounds the cut $%
\left[ w_0,+\infty \right] $, namely $\sum_{j=0}^q\beta _j>1.$ Let us
generalize the discrete sum (B7) using the integral representation 
\begin{equation}
\left( a+b\right) ^\gamma =\frac 1{\Gamma \left( -\gamma \right) }\frac
1{2i\pi }\int_{\sigma -i\infty }^{\sigma +i\infty }dz\;\Gamma \left(
-z\right) \Gamma \left( z-\gamma \right) \;a^z\;b^{\gamma -z}  \tag{B11}
\end{equation}

\noindent where $\func{Re}\gamma \leq \sigma \leq 0.$ Strictly speaking,
this integral representation is a priori valid for Re$\gamma <0.$ However,
we may, by deformation of the contour, generalize to Re$\gamma >0$ provided
that $\left| \func{Im}\gamma \right| $\TEXTsymbol{>}0 in order to avoid a
possible pinch of the contour and the extraction of residues. Consequently,
we assume that $\beta _0$ has a small imaginary part for the demonstration
and we analytically continue the result to real $\beta _0.\;$We generalize
(B11) and write (B7) as 
\begin{eqnarray}
\prod_{i=1}^{q-1}\left| \xi _i-w_0\right| ^{\beta _0-1} &=&\frac 1{\Gamma
\left( 1-\beta _0\right) }\frac 1{\left( 2i\pi \right) ^{q-1}}\int ..\int
\prod_{j=1}^{q-1}dz_j.  \nonumber \\
&&.\prod_{j=1}^q\left\{ \Gamma \left( -z_j\right)
\;A_j^{z_j}\;\prod_{i=1}^{q-1}\left| \xi _i-w_j\right| ^{z_j}\right\}
\;\delta _{Z,\beta _0-1}  \tag{B12}
\end{eqnarray}

\noindent where $\delta _{Z,\beta _0-1}$is a symbol which simplifies the
writing and which means that 
\[
z_q=\beta _0-1-\sum_{j=1}^{q-1}z_j 
\]

We may now apply (B1, B3) in order to integrate the $d\xi $ integrals with $%
\beta _j\;$replaced by $\beta _j+z_j\;$and get 
\begin{eqnarray}
&&\frac 1{\Gamma \left( \sum_{j=0}^q\beta _j-1\right) }\frac 1{\Gamma \left(
1-\beta _0\right) }\prod_{1\leq i<j\leq q}\left( w_i-w_j\right) ^{\beta
_i+\beta _j-1}.  \nonumber \\
&&\frac 1{\left( 2i\pi \right) ^{q-1}}\int ..\int
\prod_{j=1}^{q-1}dz_j\prod_{j=1}^q\left\{ \Gamma \left( -z_j\right) \Gamma
\left( \beta _j+z_j\right) \left( w_0-w_j\right) ^{\beta _0-1-z_j}\right\}
\;\delta _{Z,\beta _0-1}  \nonumber \\
&&\,  \tag{B13}
\end{eqnarray}

\bigskip\ 

\bigskip\ 

On the other hand, the integral 
\begin{equation}
\frac 1{2i\pi }\oint_{C_0}dz\frac 1{\left( w_0-z\right) ^{\beta
_0}\prod_{j=1}^q\left( z-w_j\right) ^{\beta _j}}  \tag{B14}
\end{equation}

\noindent can be calculated from the integral of the discontinuity along the
cut $\left[ w_0,+\infty \right] $ 
\begin{equation}
-\frac{\sin \pi \beta _0}\pi \int_0^\infty dx\;x^{-\beta
_0}\;\prod_{j=1}^q\left( x+w_0-w_j\right) ^{-\beta _j}  \tag{B15}
\end{equation}

\noindent Next, we use the integral representation (B11), for j=1 to q-1, in
order to write (B15) as 
\begin{eqnarray*}
&&-\frac{\sin \pi \beta _0}\pi \frac 1{\prod_{j=1}^{q-1}\Gamma \left( \beta
_j\right) }. \\
&&.\frac 1{\left( 2i\pi \right) ^{q-1}}\int ..\int \prod_{j=1}^{q-1}\left[
dz_j\Gamma \left( -z_j\right) \Gamma \left( \beta _j+z_j\right) \left(
w_0-w_j\right) ^{-z_j-\beta _j}\right] . \\
&&.\int_0^\infty dx\;x^{-\beta _0+\sum_{j=1}^{q-1}z_j}\;\left(
x+w_0-w_q\right) ^{-\beta _q}
\end{eqnarray*}

\noindent After integration of the dx integral, (B14) is equal to 
\begin{eqnarray}
&&-\frac{\sin \pi \beta _0}\pi \frac 1{\prod_{j=1}^q\Gamma \left( \beta
_j\right) }.  \tag{B16} \\
&&.\frac 1{\left( 2i\pi \right) ^{q-1}}\int ..\int
\prod_{j=1}^{q-1}dz_j\prod_{j=1}^q\left[ \Gamma \left( -z_j\right) \Gamma
\left( \beta _j+z_j\right) \left( w_0-w_j\right) ^{-z_j-\beta _j}\right]
\;\delta _{Z,\beta _0-1}  \nonumber
\end{eqnarray}

\smallskip\ 

\noindent If we compare (B16) and (B13), we just proved (B10) for a non
integer $\beta _0.$

\bigskip\ 

This ends the calculation of $K_{q-1}$ which represents the central part in
the proof of the conjecture.

\bigskip\ 

\bigskip\ 

\section{Appendix C}

\bigskip\ 

In this appendix we calculate and we give some properties of the expression 
\begin{equation}
A=\frac{\partial _{v_1}..\partial _{v_p}\left[
\prod_{i=0}^q\prod_{j=1}^p\left( w_i-v_j\right) ^{1-\beta
_i}\prod_{k<l}\left( v_k-v_l\right) \right] }{\prod_{i=0}^q\prod_{j=1}^p%
\left( w_i-v_j\right) ^{1-\beta _i}\prod_{k<l}\left( v_k-v_l\right) } 
\tag{C1}
\end{equation}

\noindent In order to calculate this expression, we simply calculate the
residues of the poles in the variables $v_j$; the first property which is
clear is that there is no singularity when two $v$'s coincide. The only way
to get a pole at$\;v_i=v_j$ is when the derivatives $\partial _{v_i}$ or/and 
$\partial _{v_j}$ act upon the term $\left( v_i-v_j\right) $ at the
numerator of (C1). Clearly, for $f(w,v)$ symmetric in $v_i$ and $v_j$, 
\begin{equation}
\frac{\partial _{v_i}\partial _{v_j}\left[ f\left( w,v\right) \left(
v_i-v_j\right) \right] }{f\left( w,v\right) \left( v_i-v_j\right) }=\frac{%
f_{ij}^{^{\prime \prime }}\left( w,v\right) }{f\left( w,v\right) }-\frac{%
f_i^{^{\prime }}\left( w,v\right) -f_j^{^{\prime }}\left( w,v\right) }{%
f\left( w,v\right) \left( v_i-v_j\right) }  \tag{C2}
\end{equation}

\noindent and the residue of the pole at $v_i=v_j$ is zero (and this
property remains evidently true when performing the other derivatives).
Consequently, we are going to extract the poles when $v_i=w_k$ neglecting
systematically all poles at $v_i=v_j$ since their residues vanish. We write 
\begin{equation}
A=\prod_{i=0}^q\frac{R_i}{w_i-v_1}  \tag{C3}
\end{equation}

\noindent and we find 
\begin{equation}
R_i=\left( \beta _i-1\right) \frac{\partial _{v_2}..\partial _{v_p}\left[
f_2..f_p\prod_{j>1}\left( w_i-v_j\right) \Delta _1\right] }{%
f_2..f_p\prod_{j>1}\left( w_i-v_j\right) \Delta _1}  \tag{C4}
\end{equation}

\noindent where $f_k=\prod_{i=0}^q\left( w_i-v_k\right) ^{1-\beta _i}\;$and $%
\Delta _1=\prod_{1<i<j}\left( v_i-v_j\right) .$

\smallskip\ 

We now extract the poles in$\;v_2$ and write 
\begin{equation}
A=\sum_{\left\{ i_1,i_2\right\} =0}^q\frac{R_{i_1,i_2}}{\left(
w_{i_1}-v_1\right) \left( w_{i_2}-v_2\right) }  \tag{C5}
\end{equation}

\noindent We find 
\begin{eqnarray}
R_{i_1,i_2} &=&\left( \beta _{i_1}-1\right) \left( \beta _{i_2}-1-\delta
_{i_{1,},i_2}\right) .  \nonumber \\
&&.\frac{\partial _{v_3}..\partial _{v_p}\left[ f_3..f_p\prod_{j>2}\left(
w_{i_1}-v_j\right) \left( w_{i_2}-v_j\right) \Delta _2\right] }{%
f_3..f_p\prod_{j>2}\left( w_{i_1}-v_j\right) \left( w_{i_2}-v_j\right)
\Delta _2}  \tag{C6}
\end{eqnarray}
\noindent where $\Delta _2=\prod_{2<i<j}\left( v_i-v_j\right) .$ We can
proceed and extract successively all the poles in $v_i$; the result can be
written in the following way: given (q+1) complementary (and possibly empty)
sets $I_0,I_1,..,I_q$ such that 
\begin{eqnarray}
I_r\cap I_s &=&\Phi  \nonumber \\
\cup _{r=0}^qI_r &=&\left( 1,..,p\right)  \tag{C7}
\end{eqnarray}

\noindent then 
\begin{equation}
A=\sum_{\left\{ I_0,..,I_q\right\} }\prod_{r=0}^q\left\{ \frac{\Gamma \left(
\beta _r\right) }{\Gamma \left( \beta _r-\left| I_r\right| \right) }%
\prod_{j\in I_r}\frac 1{w_r-v_j}\right\}  \tag{C8}
\end{equation}

\noindent where $\left| I_r\right| $ is the number of elements in $I_r.$

\smallskip\ 

Let us write, as an example, the case where $\beta =p$ integer and where we
have only two variables $w_0$ and $w_{1.}$Then, 
\begin{equation}
A=\left( p-1\right) \left( p-1\right) !\sum_{I\subset \left( 1,..,p\right)
}C_{p-2}^{\left| I\right| -1}\prod_{i\in I}\frac 1{w_0-v_i}\prod_{i\notin
I}\frac 1{w_1-v_i}  \tag{C9}
\end{equation}

\bigskip\ 

\bigskip\ 

In the rest of the appendix, we show that A is the ratio of two
determinants, and we develop a recurrent construction satisfied by
expressions of the same type of A.

\noindent We define successively the following symbols: given the symbol $%
\left[ u\right] =\left[ u\right] _0$ attached to a given variable u, and the
symbols 
\begin{equation}
\left[ u\right] _n=nu^{n-1}+u^n\left[ u\right]  \tag{C10}
\end{equation}

\noindent then, we define the completely symmetric symbols 
\begin{equation}
\left[ v_1,..,v_p\right] =\frac 1\Delta \det \left| 
\begin{array}{ccccc}
\left[ v_1\right] _0 & \left[ v_2\right] _0 & . & . & \left[ v_p\right] _0
\\ 
\left[ v_1\right] _1 & \left[ v_2\right] _1 & . & . & \left[ v_p\right] _1
\\ 
. & . &  &  & . \\ 
. & . &  &  & . \\ 
\left[ v_1\right] _{p-1} & \left[ v_2\right] _{p-1} & . & . & \left[
v_p\right] _{p-1}
\end{array}
\right|  \tag{C11}
\end{equation}

\noindent where $\Delta =\left( -\right) ^{n(n-1)/2}\prod_{i<j}\left(
v_i-v_j\right) $ is the Vandermonde determinant. Clearly, these symbols can
also be defined recursively by 
\begin{eqnarray}
\left[ v,w\right] &=&\frac{\left( 1+v\left[ v\right] \right) \left[ w\right] 
}{v-w}+\frac{\left( 1+w\left[ w\right] \right) \left[ v\right] }{w-v} 
\nonumber \\
\left[ u,v,w\right] &=&\frac{\left( 2u+u^2\left[ u\right] \right) \left[
v,w\right] }{\left( u-v\right) \left( u-w\right) }+\;circ.perm.  \nonumber \\
&&...  \nonumber \\
\left[ v_1,..,v_p\right] &=&\frac{\left( \left( p-1\right)
v_1^{p-2}+v_1^{p-1}\left[ v_1\right] \right) \left[ v_2,..,v_p\right] }{%
\prod_{i=2}^p\left( v_1-v_i\right) }+circ.perm.  \tag{C12}
\end{eqnarray}

To illustrate this construction, we give two examples. First, let us define 
\begin{equation}
\left[ u\right] =\frac \beta u  \tag{C13}
\end{equation}

\noindent then, it is easy to construct the different symbols; using the
properties of determinants, we obtain 
\begin{equation}
\left[ v_1,..,v_p\right] =\frac{\Gamma \left( \beta +p\right) }{\Gamma
\left( \beta \right) }\frac 1{v_1...v_p}  \tag{C14}
\end{equation}

\noindent Moreover, since 
\[
\left[ v\right] _n=\frac{\partial _v\left( v^{\beta +n}\right) }{v^\beta } 
\]
\noindent we have 
\begin{equation}
\left[ v_1,..,v_p\right] =\frac{\partial _{v_1}...\partial _{v_p}\left[
v_1^\beta ..v_p^\beta \prod_{i<j}\left( v_i-v_j\right) \right] }{v_1^\beta
..v_p^\beta \prod_{i<j}\left( v_i-v_j\right) }  \tag{C15}
\end{equation}

\bigskip\ 

As a second example, we define 
\begin{equation}
\left[ u\right] =\sum_{i=0}^q\frac{\beta _i-1}{w_i-u}  \tag{C16}
\end{equation}

\noindent Then, from 
\begin{equation}
\left[ v\right] _n=\frac{\partial _v\left[ v^n\prod_{i=0}^q\left(
w_i-v\right) ^{1-\beta _i}\right] }{\prod_{i=0}^q\left( w_i-v\right)
^{1-\beta _i}}  \tag{C17}
\end{equation}

\noindent and the properties of determinants, we find that 
\begin{equation}
\left[ v_1,..,v_p\right] =A  \tag{C18}
\end{equation}

\noindent as given in (C1) and (C8).

\bigskip\ \-

\bigskip\ 

\section{Appendix D}

\bigskip\ 

We are interested in the structure of the integrals 
\begin{equation}
I_{P\left( z\right) }=\frac 1{2i\pi }\oint_{C_0}dz\frac{P\left( z\right) }{%
\left( w_0-z\right) ^{\beta _0}\prod_{j=1}^q\left( z-w_j\right) ^{\beta _j}}
\tag{D1}
\end{equation}

\noindent where $P\left( z\right) $ is a polynomial and where the sum over
the variables $\beta $ is large enough to make the integral convergent along
the cut $C_0=\left[ w_0,+\infty \right] .$ To simplify the writing of this
appendix, we keep abusively a z dependence on the left hand side of (D1).
From 
\begin{equation}
\frac 1{2i\pi }\oint_{C_0}dz\;\frac \partial {\partial z}\left\{ \frac{z^n}{%
\left( w_0-z\right) ^{\beta _0-1}\prod_{j=1}^q\left( z-w_j\right) ^{\beta
_j-1}}\right\} =0  \tag{D2}
\end{equation}

\noindent we obtain the relation 
\begin{equation}
-n\;I_{z^{n-1}\prod_{j=0}^q\left( z-w_j\right) }+\sum_{j=0}^q\left( \beta
_j-1\right) I_{z^n\prod_{i\neq j}\left( z-w_i\right) }=0  \tag{D3}
\end{equation}

\noindent In the special case where n=0, we have 
\begin{equation}
\sum_{j=0}^q\left( \beta _j-1\right) I_{\prod_{i\neq j}\left( z-w_i\right)
}=0  \tag{D4}
\end{equation}

\smallskip\ 

Equation (D4) shows that the function $I_{z^q}$ is linearly dependent of the
functions $I_{z^p}$ for 0$\leq p<q.$ Also, for n\TEXTsymbol{>}0, equation
(D3) shows that the functions $I_{z^p}$ for p\TEXTsymbol{>}q can be
decomposed on the basis of the q functions $I_{z^p}$ (0$\leq p<q).$ We have
constructed a q dimensional vectorial space $E_q.$

The decomposition of the functions $I_{P\left( z\right) }$ on the basis of
functions $I_{z^p}\;(0\leq p<q)$ is particularly simple if the polynomial $%
P\left( z\right) $ is of degree \TEXTsymbol{<}q; it becomes particularly
difficult otherwise because of the relations (D3-4). We are going to
introduce other basis for the vectorial space $E_q$ which are specially
adapted when the polynomial $P\left( z\right) $ is known from its roots $%
P\left( z\right) =\prod_i\left( z-v_i\right) .$ Let us give ourselves q
different values of the variable $v$ say $v_1,..,v_q$; then, we define a
basis of $E_q$ as a set of q independent functions 
\begin{equation}
I_{\left( j\right) }=I_{\prod_{i\neq j}\left( z-v_i\right) }\;for\;j=1,..,q 
\tag{D5}
\end{equation}

\noindent The transformation from the basis $I_{z^p}$ to the basis (D5) can
be written as 
\begin{equation}
I_{z^p}=\sum_{j=1}^q\;\frac{v_j^p}{\prod_{i\neq j}\left( v_j-v_i\right) }%
\;I_{\left( j\right) }  \tag{D6}
\end{equation}

\noindent and consequently, if $P_{q-1}\left( z\right) $ is a polynomial in
z of degree at most q-1, then 
\begin{equation}
I_{P_{q-1}\left( z\right) }=\sum_{j=1}^q\;\frac{P_{q-1}\left( v_j\right) }{%
\prod_{i\neq j}\left( v_j-v_i\right) }\;I_{\left( j\right) }  \tag{D7}
\end{equation}

\bigskip\ 

As we already mentionned, the decomposition of $I_{z^q},I_{z^{q+1}}...$is
more elaborate; let us give the exemple of $I_{z^q}.$ From (D4), we have 
\begin{equation}
D_{q+1}I_{z^q}+\sum_{j=0}^q\left( \beta _j-1\right) \;I_{\prod_{i\neq
j}\left( z-w_i\right) -z^q}=0  \tag{D8}
\end{equation}

\noindent where $D_p=\sum_{j=0}^q\beta _j-p.$ Since the polynomial $%
\prod_{i\neq j}\left( z-w_i\right) -z^q$ is of degree q-1 in z, we may apply
(D7); we obtain 
\begin{equation}
I_{z^q}=\sum_{r=1}^q\frac{\left[ v_r^q-\frac 1{D_{q+1}}\sum_{j=0}^q\left(
\beta _j-1\right) \prod_{i\neq j}\left( v_r-w_i\right) \right] }{%
\prod_{s\neq r}\left( v_r-v_s\right) }\;I_{\left( r\right) }  \tag{D9}
\end{equation}

\noindent Given a polynomial $P_q\left( z\right) $ of degree q in z (with
the coefficient of $z^q$=1), we may write 
\begin{equation}
I_{P_q\left( z\right) }=\sum_{r=1}^q\frac{\left[ P_q\left( v_r\right) -\frac
1{D_{q+1}}\sum_{j=0}^q\left( \beta _j-1\right) \prod_{i\neq j}\left(
v_r-w_i\right) \right] }{\prod_{s\neq r}\left( v_r-v_s\right) }\;I_{\left(
r\right) }  \tag{D10}
\end{equation}

\noindent If moreover, the choice of the variables $v_1,..,v_q$ which
defines the basis $I_{\left( r\right) }$ coincide with the roots of the
polynomial $P_q\left( z\right) $, that is $P_q\left( v_r\right) =0$ \ for $%
\;r=1,..,q$ , then 
\begin{equation}
I_{\prod_{i=1}^q\left( z-v_i\right) }=\frac{\left( -\right) ^{q+1}}{D_{q+1}}%
\sum_{r=1}^q\frac{\prod_{i=0}^q\left( w_i-v_r\right) }{\prod_{s\neq r}\left(
v_r-v_s\right) }\;\left[ v_r\right] \;I_{\left( r\right) }\;  \tag{D11}
\end{equation}

\noindent where the symbol $\left[ v\right] $ is defined in Appendix C (C6).

\bigskip\ 

We now extend the above construction to polynomials of degree larger than q.
Given a set of p\TEXTsymbol{>}q values $v_1,..,v_p$ of a variable $v$, we
define a set of p functions as 
\begin{equation}
\overline{I_{\left( r\right) }}=I_{\prod_{i\neq r}\left( z-v_i\right) } 
\tag{D12}
\end{equation}

\noindent Clearly these p functions are necessarily dependent in the
vectorial space $E_q.$ Nevertheless, the relation (D6) remains valid 
\begin{equation}
I_{z^j}=\sum_{r=1}^p\frac{v_r^j}{\prod_{i\neq r}\left( v_r-v_i\right) }\;%
\overline{I_{\left( r\right) }}\;\;for\;j=0,1,..,p-1  \tag{D13}
\end{equation}

\noindent In order to express $I_{z^p}$, we use (D3) with n=p-q; we get 
\[
D_{p+1}I_{z^p}-\left( p-q\right) I_{z^{p-q-1}\left[ \prod_{j=0}^q\left(
z-w_j\right) -z^{q+1}\right] }+\sum_{j=0}^q\left( \beta _j-1\right)
I_{z^{p-q}\left[ \prod_{i\neq j}\left( (z-w_i\right) -z^q\right] }=0 
\]

\noindent so that, from (D13) 
\begin{eqnarray}
I_{z^p} &=&\sum_{r=1}^p\left\{ v_r^p+\frac{\left( -\right) ^{q+1}}{D_{p+1}}%
\prod_{j=0}^q\left( w_j-v_r\right) \left[ \left( p-q\right)
v_r^{p-q-1}+v_r^{p-q}\left[ v_r\right] \right] \;\right\} .  \nonumber \\
&&.\frac{\overline{I_{\left( r\right) }}}{\prod_{i\neq r}\left(
v_r-v_i\right) }  \tag{D14}
\end{eqnarray}

\noindent Now, given a polynomial $P\left( z\right) =\prod_{i=1}^p\left(
z-v_i\right) $ where the $v$'s are supposed to be all different, we may
write 
\begin{equation}
I_{\prod_{i=1}^p\left( z-v_i\right) }=\frac{\left( -\right) ^{q+1}}{D_{p+1}}%
\sum_{r=1}^p\frac{\prod_{j=0}^q\left( w_j-v_r\right) }{\prod_{i\neq r}\left(
v_r-v_i\right) }\;\left[ v_r\right] _{p-q}\;\overline{I_{\left( r\right) }}\;
\tag{D15}
\end{equation}

\noindent where $\left[ v_r\right] _n$ is defined in Appendix C (C10).

Let us illustrate this result on the example where p=q+1. In that case the
q+1 functions $\overline{I_{\left( r\right) }}$ can themselves be decomposed
according to (D11) generating a set of q(q+1)/2 functions 
\begin{equation}
\overline{I_{\left( r,s\right) }}=\prod\Sb i=1  \\ i\neq r,s  \endSb %
^{q+1}\left( z-v_i\right)  \tag{D16}
\end{equation}

\noindent When transforming (D15) in terms of $\overline{I_{\left(
r,s\right) }}$,$\;$we obtain two contributions depending whether we express $%
\overline{I_{\left( r\right) }}$ or $\overline{I_{\left( s\right) }}$ in
terms of $\overline{I_{\left( r,s\right) }}$; these two contributions
construct the combination 
\begin{equation}
\frac{\left[ v_r\right] _1\left[ v_s\right] _0}{v_r-v_s}+\frac{\left[
v_s\right] _1\left[ v_r\right] _0}{v_s-v_r}=\left[ v_r,v_s\right]  \tag{D17}
\end{equation}

\noindent where $\left[ v_r,v_s\right] $ is defined in Appendix C (C12). The
expression (D15) becomes 
\begin{equation}
I_{\prod_{i=1}^{q+1}\left( z-v_i\right) }=\frac{\left( -\right) ^{2q+2}}{%
D_{q+2}D_{q+1}}\sum_{r<s}\frac{\prod_{j=0}^q\left[ \left( w_j-v_r\right)
\left( w_j-v_s\right) \right] }{\prod_{i\neq r,s}\left[ \left(
v_r-v_i\right) \left( v_s-v_i\right) \right] }\;\left[ v_r,v_s\right] \;%
\overline{I_{\left( r,s\right) }}  \tag{D18}
\end{equation}

\noindent We insist again on the fact that this decomposition is not a
decomposition over a basis of $E_q$ but it is exactly the kind of
decomposition we need to prove the conjecture in the case q-1$\leq p.$

\bigskip\ 

We now proceed by recurrence for any p\TEXTsymbol{>}q. We suppose that for
p-1, we have 
\begin{eqnarray}
I_{\prod_{i=1}^{p-1}\left( z-v_i\right) } &=&\left( -\right) ^{\left(
p-q\right) \left( q+1\right) }\frac{\Gamma \left( \sum_{j=0}^q\beta
_j-p\right) }{\Gamma \left( \sum_{j=0}^q\beta _j-q\right) }.  \nonumber \\
&&.\sum_{J_{q-1}}\frac{\prod_{j=0}^q\prod_{k\notin J_{q-1}}\left(
w_j-v_k\right) }{\prod_{r\notin J_{q-1}}\prod_{i\in J_{q-1}}\left(
v_r-v_i\right) }\left[ \cup _{k\notin J_{q-1}}(v_k)\right] \;\overline{%
I_{J_{q-1}}}  \tag{D19}
\end{eqnarray}

\noindent where $J_{q-1}$is a set of q-1 indices $\left(
j_1,..,j_{q-1}\right) \subset \left( 1,..,p-1\right) $ and $\left[ \cup
_{k\notin J_{q-1}}\left( v_k\right) \right] $ is a notation for the symbol
of Appendix C (C11) where the indices $k$ belong to the complement of $%
J_{q-1}$ in (1,..,p-1). The notation $\overline{I_{J_{q-1}}}$ means $%
\overline{I_{\left( \cup k\notin J_{q-1}\right) }}.$ Now, we consider (D15)
and we replace $\overline{I_{\left( r\right) }}$ by its expression (D19); we
get 
\begin{eqnarray}
I_{\prod_{i=1}^p\left( z-v_i\right) } &=&\left( -\right) ^{\left(
p-q+1\right) \left( q+1\right) }\frac{\Gamma \left( \sum_{j=0}^q\beta
_j-p-1\right) }{\Gamma \left( \sum_{j=0}^q\beta _j-q\right) }\sum_{r=1}^p%
\frac{\prod_{j=0}^q\left( w_j-v_r\right) }{\prod_{i\neq r}\left(
v_r-v_i\right) }\;\left[ v_r\right] _{p-q}.  \nonumber \\
&&.\sum_{J_{q-1}}\frac{\prod_{j=0}^q\prod\Sb k\notin J_{q-1}  \\ k\neq r 
\endSb \left( w_j-v_k\right) }{\prod\Sb s\notin J_{q-1}  \\ s\neq r  \endSb %
\prod_{i\in J_{q-1}}\left( v_s-v_i\right) }\;\left[ \cup \Sb k\notin J_{q-1} 
\\ k\neq r  \endSb \left( v_k\right) \right] \;\overline{I_{J_{q-1}}} 
\tag{D20}
\end{eqnarray}

\noindent where $J_{q-1}=\left( j_1,..,j_{q-1}\right) \subset \left\{ \left(
1,..,p\right) -\left\{ r\right\} \right\} .$ We now invert the sum over r
and the sum over $J_{q-1,}$%
\begin{eqnarray}
I_{\prod_{i=1}^p\left( z-v_i\right) } &=&\left( -\right) ^{\left(
p-q+1\right) \left( q+1\right) }\frac{\Gamma \left( \sum_{j=0}^q\beta
_j-p-1\right) }{\Gamma \left( \sum_{j=0}^q\beta _j-q\right) }.  \nonumber \\
&&.\sum_{J_{q-1}}\frac{\prod_{j=0}^q\prod_{k\notin J_{q-1}}\left(
w_j-v_k\right) }{\prod_{s\notin J_{q-1}}\prod_{i\in J_{q-1}}\left(
v_s-v_i\right) }.  \nonumber \\
&&.\sum_{r\notin J_{q-1}}\frac{\left[ v_r\right] _{p-q}\;\left[ \cup \Sb %
k\notin J_{q-1}  \\ k\neq r  \endSb \left( v_k\right) \right] }{\prod\Sb %
i\notin J_{q-1}  \\ i\neq r  \endSb \left( v_r-v_i\right) }\;\;\;\overline{%
I_{J_{q-1}}}  \tag{D21}
\end{eqnarray}

\noindent where $J_{q-1}=\left( j_1,..,j_{q-1}\right) \subset \left(
1,..,p\right) .$By Appendix C (C12), the sum over $r$ in the second line of
(D21) is nothing but $\left[ \cup _{k\notin J_{q-1}}\left( v_k\right)
\right] .$ We have obtained (D19) where p has been replaced by p+1; this
ends the proof of the recurrence.

\bigskip\ 

\bigskip\ 

We now rewrite (D21) in a way which is directly applicable to section 2
eq.(2.20-21). We have 
\begin{eqnarray}
&&\frac 1{\prod_{s\notin J_{q-1}}\prod_{i\in J_{q-1}}\left( v_s-v_i\right) }
\nonumber \\
&=&\left( -\right) ^{(2p-q)\left( q-1\right) /2+\delta _{J_{q-1}}}\;\frac{%
\prod\Sb k<l  \\ k,l\notin J_{q-1}  \endSb \left( v_k-v_l\right) \;\prod\Sb %
k<l  \\ k,l\in J_{q-1}  \endSb \left( v_k-v_l\right) }{\prod_{k<l}\left(
v_k-v_l\right) }  \tag{D22}
\end{eqnarray}

\noindent where $\delta _{J_{q-1}}=\sum_{k=1}^{q-1}j_k-q+1.$ From the
definition of the symbol $\left[ \cup _{k\notin J_{q-1}}\left( v_k\right)
\right] $ as given in Appendix C (C18,C1), we may write 
\begin{eqnarray}
&&\sum_{J_{q-1}}\left( -\right) ^{\delta _{J_{q-1}}}\frac{\prod_{j\notin
J_{q-1}}\left( \partial _{v_j}\right) \left[ \prod_{i=0}^q\prod_{j\notin
J_{q-1}}\left( w_i-v_j\right) ^{1-\beta _i}\prod\Sb k<l  \\ k,l\notin
J_{q-1}  \endSb \left( v_k-v_l\right) \right] }{\prod_{i=0}^q\prod_{j\notin
J_{q-1}}\left( w_i-v_j\right) ^{-\beta _i}}.  \nonumber \\
&&.\frac{\prod\Sb k<l  \\ k,l\in J_{q-1}  \endSb \left( v_k-v_l\right) }{%
2i\pi }\oint_{C_0}dz\frac{\prod_{j\in J_{q-1}}\left( z-v_j\right) }{\left(
w_0-z\right) ^{\beta _0}\prod_{j=1}^q\left( z-w_j\right) ^{\beta _j}} 
\nonumber \\
&=&\left( -\right) ^{\left( q-1\right) \left( q-2\right) /2}\;\;\frac{\Gamma
\left( \sum_{j=0}^q\beta _j-q\right) }{\Gamma \left( \sum_{j=0}^q\beta
_j-p-1\right) }.  \nonumber \\
&&\frac{\prod_{k<l}\left( v_k-v_l\right) }{2i\pi }\oint_{C_0}dz\frac{%
\prod_{j=1}^p\left( z-v_j\right) }{\left( w_0-z\right) ^{\beta
_0}\prod_{j=1}^q\left( z-w_j\right) ^{\beta _j}}  \tag{D23}
\end{eqnarray}

\noindent which is the relation used at the end of section 2.

\newpage\ 

\noindent $\;1$ F.Calogero, J.Math.Phys.\textbf{10, }2191, 2197\textbf{\ (}%
1969\textbf{),}

\noindent \ \ \ B.Sutherland, J.Math.Phys. \textbf{12}, 246 (1971);
Phys.Rev. \textbf{A4}, 2019 (1971);

\noindent  \ \ \ \textbf{5}, 1372 (1972).

\noindent $\;2$ P.J.Forrester, Nucl.Phys. \textbf{B388}, 671 (1992)

\noindent $\;3$ F.J.Dyson, J.Math.Phys.\textbf{\ 3}, 140, 157 (1962),

\noindent \thinspace \thinspace \thinspace \thinspace \thinspace \thinspace
\thinspace B.D.Simons, P.A.Lee, B.L.Al'tshuler, Nucl.Phys. \textbf{B409},
487 (1993),

\noindent  $\;\,\,\,\,\,$K.B.Efetov, Adv.Phys. \textbf{32}, 53 (1983).

\noindent $\,\,4$ F.D.M.Haldane, M.R.Zirnbauer, Phys.Rev.Lett. \textbf{71},
4055 (1993),\ 

\noindent \ \ \ M.R.Zirnbauer, F.D.M.Haldane, Phys.Rev. \textbf{B52}, 8729
(1995).

\noindent $\;5$\ F.Lesage, V.Pasquier, D.Serban, Nucl.Phys. \textbf{B435},
585 (1995).

\noindent $\,\;6$ Z.N.C.Ha, Phys.Rev.Lett. \textbf{73}, 1574 (1994);
Nucl.Phys. \textbf{B435}, 604 (1995).

\noindent $\,\;7$ D.Serban, F.Lesage, V.Pasquier, Nucl.Phys. \textbf{B466},
499 (1996).

\noindent $\,\;8$ D.V.Khveshchenko, Int.J.Mod.Phys. \textbf{B9}, 1639 (1995).

\noindent $\,\;9$ K.Mimachi, Y.Yamada, Comm.Math.Phys. \textbf{174}, 447
(1995),

\noindent \ \ \ \ H.Awata, Y.Matsuo, S.Odake, J.Shiraishi, Nucl.Phys. 
\textbf{B449}, 347 (1995).

\noindent $10$ S.N.M.Ruijsenaars, H.Schneider, Ann.Phys. \textbf{170}, 370
(1986),

\noindent \ \ \ \ S.N.M.Ruijsenaars, Comm.Math.Phys. \textbf{110}, 191
(1987),

\noindent \ \ \ \ H.Konno, Nucl.Phys. \textbf{B473}, 579 (1996).

\noindent $11$ I.G.Macdonald, Symmetric functions and Hall polynomials, 2nd
ed. (Claren

\noindent  \ \ \ \ don Press, 1995).

\noindent $12$ A.L.Dixon, Proc.London Math.Soc. (2), \textbf{3}, 206 (1905)

\noindent $13$ H.Bateman, Higher Transcendental functions, Vol \textbf{1},
p.85, eq.(13),

\noindent \ \ \ \ E.B.Elliot, Messenger of Math, \textbf{33}, 31 (1904).\ \ 

\end{document}